\def\year{2019}\relax
\documentclass[letterpaper]{article} 
\usepackage{aaai19}  
\usepackage{times}  
\usepackage{helvet}  
\usepackage{courier}  
\usepackage{url}  
\usepackage{graphicx}  

\usepackage{float}
\usepackage{booktabs} 
\usepackage{footnote}
\usepackage{amsmath}
\usepackage{amsfonts}
\usepackage{ upgreek }
\makesavenoteenv{tabular}
\makesavenoteenv{table}

\usepackage{color}

\title{DeepTileBars: Visualizing Term Distribution for Neural Information Retrieval}

\author{Zhiwen Tang
 , 
  Grace Hui Yang \\ 
  InfoSense, Department of Computer Science\\
 Georgetown University\\
 zt79@georgetown.edu,huiyang@cs.georgetown.edu
}
\begin{document}
\maketitle

\begin{abstract}
Most neural Information Retrieval (Neu-IR) models derive query-to-document ranking scores based on term-level matching. Inspired by TileBars, a classical term distribution visualization method,
in this paper, we propose a novel Neu-IR model that handles query-to-document matching at the subtopic and higher levels. Our system first splits the documents into topical segments, ``visualizes" the matchings between the query and the segments, and then feeds an interaction matrix into a Neu-IR model, DeepTileBars, to obtain the final ranking scores. DeepTileBars models the relevance signals occurring at different granularities in a document's topic hierarchy. It better captures the discourse structure of a document and thus the matching patterns. Although its design and implementation are light-weight, DeepTileBars outperforms other state-of-the-art Neu-IR models on benchmark datasets including the Text REtrieval Conference (TREC) 2010-2012 Web Tracks and LETOR 4.0.
 
\end{abstract}



\section{Introduction}


Numerous efforts have  been devoted  to advance Information Retrieval (IR) with deep neural networks \cite{Guo:2016:DRM:2983323.2983769,pang2017deeprank,hui2017pacrr,fan2017learning,pang2016text,mitra2017learning,Xiong:2017:ENA:3077136.3080809,huang2013learning,shen2014learning}. These neural Information Retrieval (Neu-IR) models are often combined with a learning-to-rank framework  \cite{liu2009learning} to derive document relevance scores. Early experiments \cite{Nguyen:2017:DDN:3121050.3121063,pang2016study} reported only marginal gains or even inferior performance to traditional IR methods such as BM25 \cite{robertson2009probabilistic} and language modeling \cite{zhai2017study}. The more recent Neu-IR models attempted to incorporate well-known information retrieval principles and have started to show improvements over traditional IR methods. 


The state-of-the-art Neu-IR models can be grouped into two categories \cite{Guo:2016:DRM:2983323.2983769}. The first category is representation-focused. These models  map the texts into a low-dimensional space and then compute document relevance scores in that space. 
Models in this family include DSSM \cite{huang2013learning} and CDSSM \cite{shen2014learning}. 
The second category is interaction-focused. They first produce an interaction matrix, a.k.a. a matching matrix, between a query 
and a document. Each entry in the matrix is usually a simple relevance measurement, such as the cosine similarity, between the query  term vector and the document term vector. The matrix is then fed into a deep neural network to learn the document relevance  score.
Models in this family include PACRR \cite{hui2017pacrr}, DeepRank \cite{pang2017deeprank}, DRMM \cite{Guo:2016:DRM:2983323.2983769},  K-NRM \cite{Xiong:2017:ENA:3077136.3080809}, MatchPyramid \cite{pang2016text} and HiNT \cite{hint2018}. There are also hybrid models, such as DUET \cite{mitra2017learning}, that combine the  scores generated by the two categories.

The interaction-focused models are more popular. Partly, it is because of their close connection to the widely used learning-to-rank method. Also, they benefit from the interaction matrix's ability to  reveal relevance signals visually, just like what a query highlighting function can do.   For  human readers, visualization could help accelerate the processing of information \cite{byrd1999scrollbar,hornbaek2001reading}. In real-world search engine development, there is also a high demand for visualization functions, such as query term highlighting and thumbnail images \cite{hearst2009search}. These visualizations  offer efficient, direct, and informative feedback to a search engine user. 
We think what makes visualization practically valuable to
humans might also be valuable to a deep neural network for its resemblance to human neurons.

\begin{figure*}[t]
\centering 
\includegraphics[width=\textwidth]{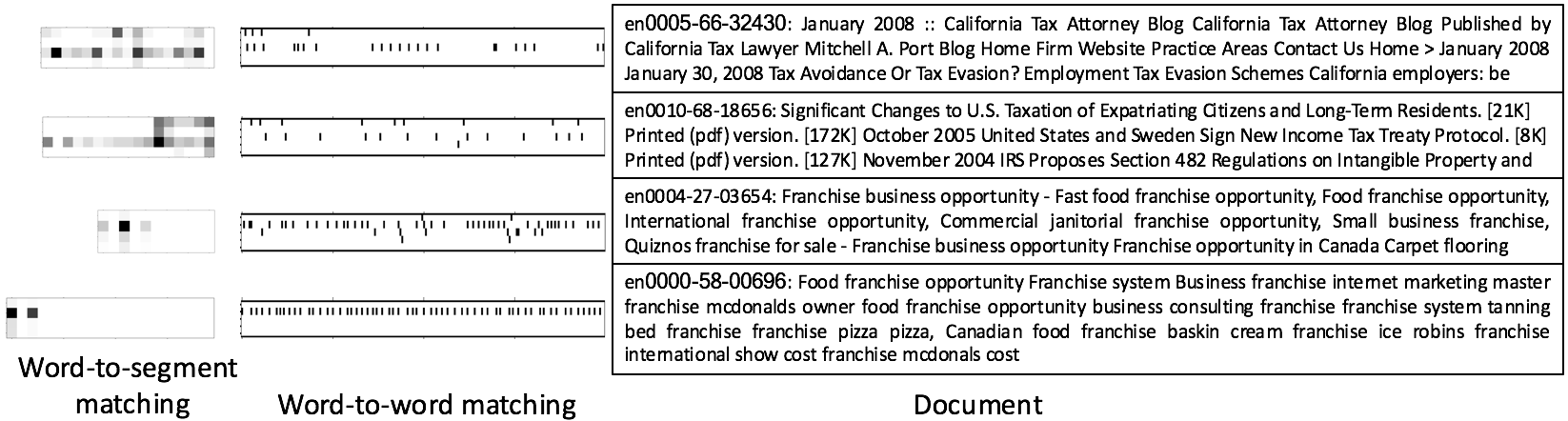}
\caption{Word-to-segment vs. Word-to-word matching. Query terms are aligned vertically and document words/segments are aligned horizontally. 
The top two documents are relevant and the bottom two are irrelevant for TREC 2011 Web Track query 116 ``California franchise tax board''.} 
\label{fig:rel_vs_irrel} 
\end{figure*}

In the history of IR, visualizing relevance signals is not a new idea. In the 1990s, Hearst proposed TileBars \cite{tilebars} to help  users better understand ad-hoc retrieval and the role played by each query term. TileBars visualizes  how query terms distribute within a document and produce a compact image to explicitly show the relevance (the matches) to the user. 
 In TileBars, a document is segmented with an algorithm called TextTiling \cite{texttiling}.  TextTiling  splits the document  at positions where  a change in topic is detected. 
 After that,  query term hits are computed per term per segment  and the distribution is plotted on a two-dimensional grid. 
 
Given a query, TileBars attempts to capture the relevance patterns in the discourse structure of a document.  Common patterns include
chained, ringed, monolith, piecewise and hierarchical \cite{Skorochodko1971,texttiling}. 
TileBars focuses on the  coarse  patterns, especially the piecewise pattern. Other work in empirical discourse processing, for instance the Rhetorical Structure Theory, uses hierarchical
 models \cite{mann1988rhetorical}. Nonetheless, most work employs segments or subtopics as the analysis unit in their discourse models.

Unlike the discourse models, most Neu-IR models use words as the analysis units. In this research, we decide to learn from the discourse models. Inspired by TileBars, this paper  studies the query-to-document matching at the segment or higher levels. 
We  propose DeepTileBars, a new deep neural network architecture that  leverages TileBars visualizations for ad-hoc text  retrieval.  
By employing semantically more meaningful matching units, i.e. the segments,  DeepTileBars is a step forward towards explainable artificial intelligence (XAI).


Figures \ref{fig:rel_vs_irrel}  illustrate four TileBars visualizations for the query `California franchise tax board'. The top two documents are relevant and the bottom two are irrelevant.  The darker the cell, the more  matches in the cell. With word-level query-to-document matching,  we can see that the matched words are scattered around. In this example,  an irrelevant document also has many dark cells because `franchise' appears frequently in a long spam passage. It is quite difficult to tell the relevant matching patterns from the irrelevant ones.  On the contrary,  with word-to-segment matching, the words that belong to the same segment collapse into a single cell. Since the spam passage  only contributes to one cell in the visualization, it is less likely for an irrelevant document to demonstrate patterns that are expected  in a relevant one -- for instance, continuous matched segments. 

The piecewise discourse pattern used  in the original TileBars paper  already 
seems to  be quite effective. We adopt it in this paper.  In addition, we go beyond TileBars   and include the hierarchical patterns in this work. To do so, 
we  utilize a bag of Convocational Neural Networks (CNNs) \cite{krizhevsky2012imagenet}, each of which makes use of  a kernel with a different size,  to model the topical structure at various granularities. Practically, our model provides a hierarchical matching for a query-document pair. A relevance pattern could appear at any granularity in the hierarchy and our model aims to capture it at any level. 

Our work focuses on text retrieval. Even though it is a visualization-based approach, image retrieval is not within our scope.  Experiments  on the Text REtrieval Conference (TREC) 2010-2012 Web Tracks \cite{trec2012} 
		and LETOR 4.0  \cite{letor} 
		show that our model outperforms the state-of-the-art text-based Neu-IR models.  

In the remainder of this paper, we first present our version of document segmentation and term distribution visualization. Note that this process belongs to the indexing phase of a search engine and will only be performed once for the entire corpus. Next, we describe an end-to-end Neu-IR model that makes use of these visualizations to retrieve documents. We then report the experiments, followed by the related work and the conclusion.

\section{Visualizing Matched Text at the Segment Level} \label{sec:tilebars}


The construction of TileBars include three parts:  segmentation, dimension standardization and  coloring. The segmentation is done by  TextTiling. 
We then prepare the interaction matrix into fixed  dimensions and `color' each cell. 

\subsection{Segmentation by TextTiling}




TextTiling is a query-independent segmentation algorithm. It assumes that each document is composed of a linear sequence of  segments, each of which represents a coherent topic. TextTiling inputs a document and outputs a list of topical segments. Figure \ref{fig:texttiling} shows an excerpt from a long document that is used as  a walking example to illustrate the algorithm.
The algorithm takes three steps: Token Sequence Generation, Similarity Computation, and Boundary Determination. 

\begin{enumerate}
	\item \textbf{Token Sequence Generation:} 
	We start with splitting a document into token sequences. A {\it token sequence} is like a pseudo-sentence, which contains a fixed number of words. 
	After removing the stopwords, every $\alpha$ (set to 
	20 as recommended by \citeauthor{texttiling}) words   are grouped into a token sequence.\footnote{In our implementation, to preserve the natural paragraph boundaries, rarely some token sequences are permitted to have  sizes slightly different from 20.} These non-overlapping token sequences are the basic units to form a segment. 
	In Figure \ref{fig:texttiling}, each line is a token sequence, denoted as $T_i$, with $i$ indexes the sequences starting from 1. 
	
	\item \textbf{Similarity Computation:} 
 Once obtaining the token sequences, the topical boundaries would be determined based on the similarity between the  sequences as well as its contexts. The context for a token sequence is a sliding window of size $\beta$.  $\beta$ is set to  6  as recommended by \citeauthor{texttiling}. 
The similarity  for two neighboring sequences is calculated over the two windows to which they each belong. Formally, given $n$ token sequences, $T_1, T_2..., T_i, ..., T_n$, the similarity between $T_i$ and $T_{i+1}$ is calculated as the cosine similarity of term counts ($tf$) in the two windows $\left[ T_{i-\beta+1},..., T_i \right]$ and $ \left[ T_{i+1},..., T_{i+\beta} \right]$:
	\begin{equation} \scriptsize
	\begin{split}
		&sim(i, i+1) = \\
		&\frac{ \sum_{w_j}   tf\big( w_j, \left[ T_{i-\beta+1},..., T_i \right]\big)  \cdot tf\big(w_j, \left[ T_{i+1},..., T_{i+\beta} \right]\big) }
		{ \sqrt{\big(\sum_{w_j}   tf\big(w_j, \left[ T_{i-\beta+1},..., T_i \right]\big)^2\big) \cdot \big(\sum_{w_j} tf\big(w_j, \left[ T_{i+1},..., T_{i+\beta} \right]\big)^2\big)}}
	\end{split}
	\end{equation}
	where $w_j$ is the $j^{th}$ word in the vocabulary,  $tf\big(w_j, \left[ T_{i-\beta+1},..., T_i \right]\big)$ is the term frequency of $w_j$ in a window holding sequences $T_{i-\beta+1}, ... ,$ up to $T_i$, and $tf\big(w_j, \left[ T_{i+1},..., T_{i+\beta} \right]\big)$ is the term frequency of $w_j$ in a window holding  $T_{i+1}, ... ,$ up to $T_{i+\beta}$. 
	
	
	Figure \ref{fig:texttiling} illustrates an example to compute  $sim(4, 5)$ with $\beta=3$. The similarity between token sequences $T_4$ and $T_5$ is computed based on the window upto and including $T_4$,  i.e. $[T_2, T_3, T_4]$, and the window after $T_4$,  i.e. $[T_5, T_6, T_7]$. We obtain  a similarity score for every neighboring pairs of token sequences. 
	

\begin{figure}[t]
\centering
\includegraphics[width=\linewidth]{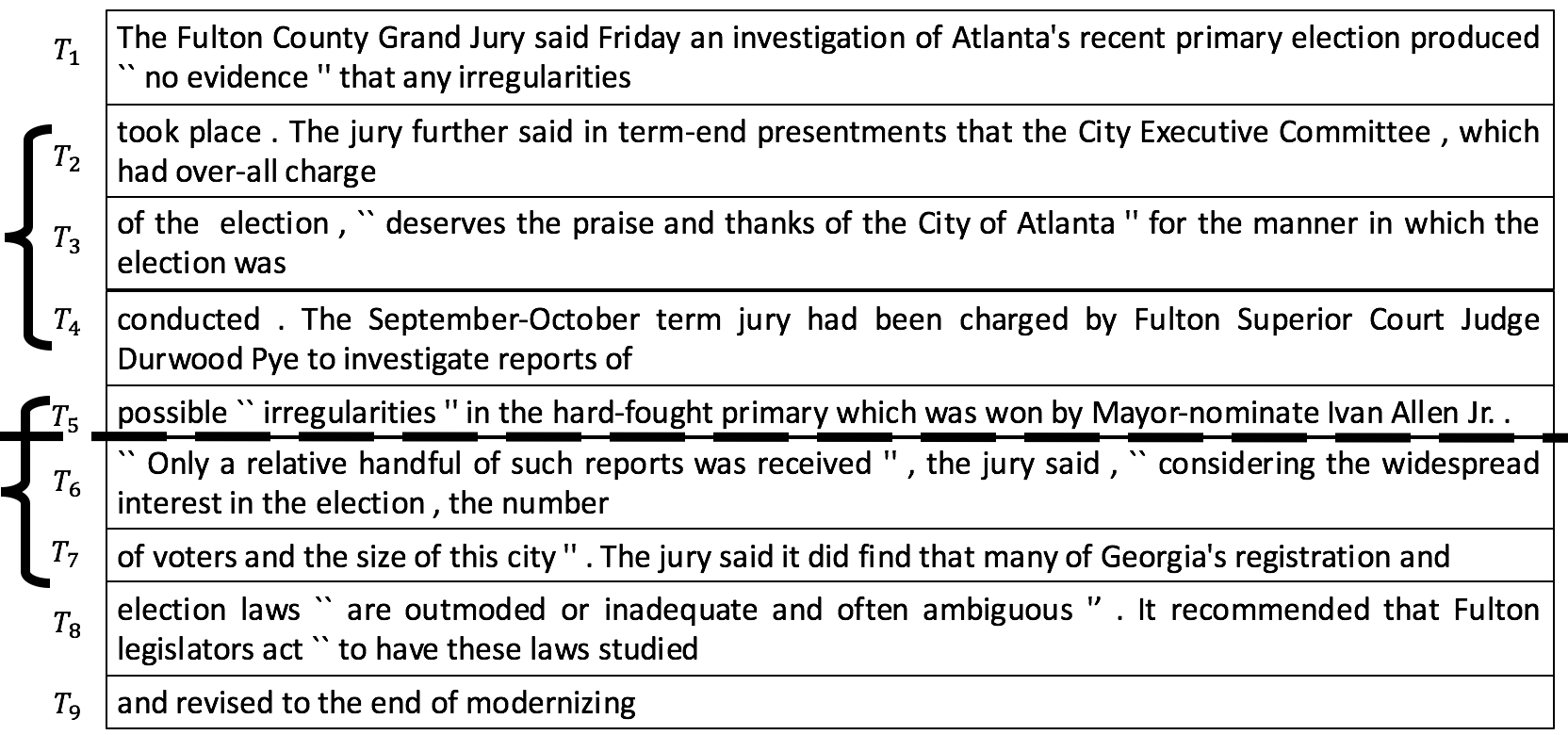}
\caption{TextTiling ($\alpha=20, \beta=3$, stopwords are kept to preserve readability).} 
\label{fig:texttiling} 
\end{figure}

	\item \textbf{Boundary Determination:} The previous steps result in a series of similarity scores, each of which is created for a pair of token sequences. 
	When plotting the similarity scores, we can observe ``peaks'' and ``valleys" in this curve. When there is a dramatic drop in the curve, we say there is a possible topic change.  
	A \textit{depth} score is computed for each neighboring pairs of token sequences. It is defined as the sum of the absolute differences to its closest neighboring ``peaks''. The {\it depth} score  between $T_i$ and $T_{i+1}$ is
	\begin{equation} \small
	\begin{split}
	depth(i, i+1) &= \mid  sim(LeftPeak(i, i+1)) - sim(i, i+1) \mid  \\
	& + \mid sim(RightPeak(i, i+1)) - sim(i, i+1) \mid 
	\end{split}
	\end{equation}
where $LeftPeak$ computes the position of the closest ``peak'' on the left side and $RightPeak$ computes that on the right side. Boundaries 
	are discovered at positions where the depth score is greater than $\mu-\updelta/2$, where $\mu$ is the mean depth  in the document and $\updelta$ is the standard deviation. The threshold adjusts to different documents. 

	
	
	In Figure \ref{fig:texttiling}, a boundary is found between $T_5$ and $T_6$. Two topical segments $B_1$ (containing $T_1, T_2, ... $) and $B_2$ (containing $T_6, T_7, T_8$ and the rest) are generated for the document. When examining the actual content of these segments, we  find  that the two segments represent distinct topics: one is the overview of an investigation and another is about the jury's opinion. 
\end{enumerate}

TextTiling was a pioneering work in the 1990s for  text segmentation. 
Unlike segmentation by a fixed passage length or by natural paragraphs, TextTiling's segments are based on term distributions and are expected to be topic-coherent. 
More sophisticated text segmentation methods  have been proposed since. Examples include probabilistic models that use language modeling and language cues \cite{beeferman1999statistical},  clustering-based algorithms \cite{kazantseva2011linear}, topic modeling \cite{misra2009text}, and the recently proposed deep learning methods \cite{badjatiya2018attention}. Compared with TextTiling, most successor approaches still hold the  assumption that topics are laid  sequentially in a document. Without loss of generality, we  use TextTiling  for its simplicity and effectiveness and focus on  how a segment-based visualization can aid  neural information retrieval. 



\subsection{Standardizing  Dimensions}

Each segment  $B_i$ would contain a  different number of token sequences; and the total number of segments in a document would also vary from document to document. Note that this number is not only decided by the choices of token-sequence length ($\alpha$) and window size ($\beta$) but also determined by the content of a  document. 
In our experiments, there are documents consisting only of a title, which yield only one segment; whereas other longer documents split into hundreds of segments. 
 As a result, for different query and document pairs,  the dimensions of  their TileBars visualizations vary.  However, a neural network requires all its input vectors to be of equal size. We therefore need a mechanism to standardize the  TileBars visualizations. 
 
 Suppose the original dimension of an image is $x \times y$, where $x$ is the query length and $y$ is the number of segments. Our task is to resize all visualizations to a chosen dimension $n_{q} \times n_{b}$.




Generally speaking, for visualizations with fewer than $n_{q}$ query terms or $n_{b}$ segments, we pad the grid with empty cells. This is equivalent to the zero padding technique  widely used in computer vision \cite{szeliski2010computer}. For query $q$, its $i^{th}$ word $w_i$ is transformed into  
\begin{equation}
w'_i = \left \{
 				\begin{tabular}{cc}
 				$w_i$ & $i \leq x$ \\
 				$\langle \mbox{empty word} \rangle $  & $ x < i \leq n_{q}$ 
 				\end{tabular}
		\right.
\end{equation}
where  $x$ is the length of original query and $n_{q}$ is the length of the transformed query and set to the maximum  query length in the query collection.  



Similarly to the query dimension, 
if $y \leq n_{b}$, i.e. the original document is relatively short, then, zero padding is done and each segment is transformed by:
\begin{equation}
B'_i = 	\left \{
 				\begin{tabular}{cc}
 				$B_i$ & $i \leq y$ \\
 				$\langle \mbox{empty segment} \rangle $ & $ y < i \leq n_{b}$ 
 				\end{tabular}
		\right.
\end{equation}
where $B'_i$ is the segment after transformation and $B_i$ is the original segment. $y$ is the original number of segments.

For  documents with more than $n_{b}$  segments, we  `squeeze'  the content after (and including) the ${n_{b}}^{th}$ segment into the ${n_{b}}^{th}$ segment, which makes the ${n_{b}}^{th}$ segment the last in a document. This would preserve all the original content while only affect ``resolution'' of the last segment.  Thus, if $y > n_{b}$, i.e., the original document is relatively long, then
\begin{equation}
B'_i = 	\left \{
 				\begin{tabular}{cc}
 				$B_i$ & $i < n_{b}$ \\
 				$concatenate(B_{n_b}, B_{n_b+1}, ... B_{y}) $  & $i=n_{b}$ 
 				\end{tabular}
		\right.
\end{equation}

 After  resizing, the visualization  dimensions are fixed as $n_{q} \times n_{b}$ for all query-document pairs.   
From  now on, we denote   query words as $w$ and  segments  as $B$ after the standardization for the sake of notation simplicity.

\subsection{Coloring}

%


The original TileBars dyes the grids based only term frequency. The darker the color, the larger the intensity. In our work, we propose to incorporate multiple relevant features, similar to  canonical colors in a color space, to ``paint" the cells.  Following the convention in multimedia research, we call each relevance feature a channel. 

In theory, features indicating query-document relevance that are effective in existing learning-to-rank models can all be used here. However, to avoid interfering with the neural network's ability to select the best feature combinations, we propose to only employ features that are independent of each other. It is similar to only use  a few scalar valued colors as the canonical   colors  in a color space, and leave the feature aggregation to the network itself. 

Three features are used in this work. They are all proven to be essential in ad-hoc retrieval. They are term frequency, inverse document frequency, and word similarity based on distributional hypothesis. 
The $(i,j)^{th}$ cell in the matching visualization $I$ for query $w$ and segment $B_j$ is `painted' by: 
\begin{equation}\label{eq:defn} \small
 \left(
\begin{array}{c}
tf(w_i, B_j)  \\
idf(w_i) \times \mathbb{I}_{B_j}\left(w_i\right)\\ 
\max_{t \in B_j} e^{-(v_{w_i} - v_t)^2}
\end{array}
\right)
\end{equation}
where $w_i$ is the $i^{th}$ query term, $B_j$ is the $j^{th}$ text segment in a document,  $tf(w_i, B_j)$ is the term frequency of $w_i$ in $B_j$, and $idf(w_i)$ is the inverse document frequency of $w_i$. $\mathbb{I}_{B_j}\left(w_i\right)$ is an indicator function to show whether $w_i$ is present in $B_j$. $v_t$ is the embedded word vector of word $t$, which comes from a pre-trained word2vec model \cite{NIPS2013_5021}.   We use Gaussian kernels  as suggested by   \cite{Xiong:2017:ENA:3077136.3080809,pang2016study}. 
The $max$ operator is used  to select the most similar word.

\section{DeepTileBar: Deep Learning with TileBars} \label{sec:neural}

We propose a novel deep neural network,  DeepTileBars, to evaluate document relevance. It consists of three layers of networks. It starts with  
detecting relevance signals with  a layer of CNNs, followed by a layer of Long Short Term Memories (LSTMs) \cite{hochreiter1997long} to aggregate the signals. And then it decides the final relevance with a Multiple Layer Perceptron (MLP).  Figure \ref{fig:network} illustrates the architecture of DeepTileBars. The  input to the network is an $n_q \times n_b$ interaction matrix. 




\subsection{Word-to-Segment Matching}




Following TileBars, DeepTileBars builds an interaction matrix by comparing a word and a topical segment. 
Each segment presumably corresponds to a topic.  
Per word information within the same segment  is absorbed into a single cell and the word-level matching is no longer supported.


Using the proposed word-to-segment matching,  we are able to produce relevance signals at the topic level and focus on the presence of query terms in consecutive topics. Matching queries to documents at the topic level, rather than at the word level,  is perhaps  easier to detect the stronger relevance signals with a bigger chunk  of coherent text.

Our design is especially beneficial for proximity queries within a large window size. 
Most existing Neu-IR models would be sufficient for proximity queries within a tight window, e.g. 2 or 3 words apart \cite{hui2017pacrr,pang2016study}. 
However, when the required window size is large, scanning a word-to-word matching with CNNs might miss the true relevant. 
A word-to-segment matching, instead, could resolve this issue by merging topically coherent words into a single segment and obtain a stronger relevance signal. 
In addition, as demonstrated in Figure \ref{fig:rel_vs_irrel}, segment-based matching would eliminate high hits in spam passages thus avoid false positives.

\subsection{Bagging with Different Kernel Sizes}

TextTiling partitions a document into segments and the segments are laid out sequentially without any further organization. However, topics in  documents are often organized hierarchically. We can imagine that the segments form sections, and section form chapters, etc. If we could put  several topical segments together, they might  actually form a super topic -- a topic that is at a higher level.  It would thus  be desirable if the topics and super topics  can be handled at  different granularities so that their hierarchical nature can be  preserved.

\begin{figure}
\includegraphics[width=0.8\linewidth]{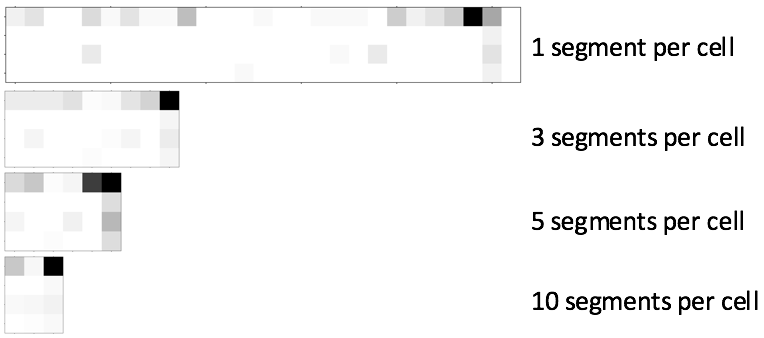}
\caption{Word-to-segment matching at different granularity levels. Document  enwp00-69-13554 for TREC 2011 Web Track query 116: California franchise tax board.}
\label{fig:topic_granularity} 
\end{figure}

Figure \ref{fig:topic_granularity} shows the word-to-segment matching at different granularity levels. We combine $k$ adjacent non-overlapping topical segments to show the term distribution at at different  levels. $k$ is the CNN kernel size. The bigger $k$ is, the larger the CNN kernel, and the higher the topic level. We can see that each granularity level provides different relevance patterns for the same document. A relevant pattern can appear at any level. Looking at the relevance patterns at all levels would thus increase our chances of finding the true relevant.

We thus propose to employ  multiple CNNs \cite{krizhevsky2012imagenet} with various kernel sizes to detect the relevance signals. Here we 
assume the kernel sizes  vary from 1 to $l$, where  $l$ is the maximum number of segments the CNNs can handle. Practically, our model provides  a  hierarchical organization:  token sequences are built from terms,  segments (topics)  are built from  token sequences,  and super topics are built by grouping adjacent topics with larger kernel sizes.

\subsection{The Network}

\begin{figure}[t] \centering
\includegraphics[width=1.05\linewidth]{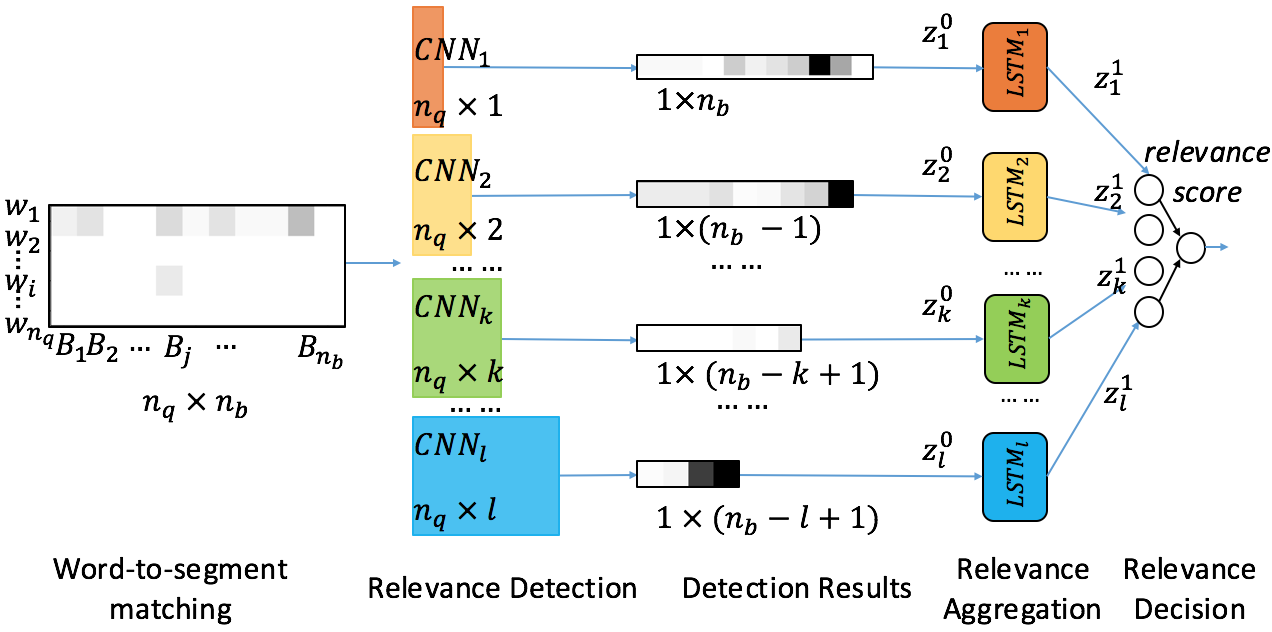}
\caption{DeepTileBars Architecture.}  
\label{fig:network}  
\end{figure}

Figure \ref{fig:network} illustrates the network architecture. 
In the first CNN layer,  
 there are $l$ CNNs, $[CNN_1, CNN_2, ... , CNN_k, ... , CNN_l ]$.  Each CNN's kernel size differs. For the $k^{th}$ CNN, its kernel size is $n_q \times k$.  
When each CNN  scans through the input visualization, it produces   a ``tape''-like output.  As $k$ increases, the CNNs are able to capture the relevance signals at $k$ adjacent text segments. We call these $k$ adjacent text segments  `$k$-segment'. 
 The $l$ number of CNNs produce $l$ outputs.  The output from the $k^{th}$ CNN denotes the relevance  detected from all $k$-segments in the document. The shape of the $k^{th}$ output is  not a square, but a thin tape with dimension $1 \times (n_b - k +1)$.   

Note that Figure \ref{fig:topic_granularity} is not the outputs of the CNNs. Instead, it shows the relevance signals per query term when combining adjacent topical segments.  The CNN outputs are the third column of thin-tapes in Figure \ref{fig:network}. 
 

The  CNN layer is denoted as $z^{0}$. The computation of $z^{0}_k$ with  kernel size $k$ is defined as 
\begin{equation}
z^{0}_k = CNN_k(I), \quad k = 1,2,3,...,l
\end{equation}
More specifically,
\begin{equation}  \small
z^{0}_k [i] = act \left( \sum_{ch=1}^{|channels|} \sum_{u=0}^{n_q-1} \sum_{v=0}^{k-1} \theta_{k,ch}^0 [u,v] * I[u, i+v, ch] + b_k^0 \right)
\end{equation} 
where $act$ is the ReLU activation function, $ch$ is the channel index,  $\theta$ is the kernel coefficient, and $b$ is the bias. 

At the second LTSM layer, in order to minimize the loss of information, we  use the same $l$ number of  LSTMs  to accumulate the relevance signals. The output of the $k^{th}$ CNN is fed into the $k^{th}$ LSTM. An LSTM scans through its input step by step from the beginning to the end.   The $k^{th}$ LSTM then outputs its evaluation of the entire document at the granularity of $k$. For instances, the first LSTM, $LSTM_1$, accumulates relevance signals from all the $1$-segments;  the second LSTM, $LSTM_2$, accumulates those from all $2$-segments.  Thus, 
 the LSTMs layer is able to output multiple relevance estimations at different  granularities.  

When scanning at the $t^{th}$ timestep, 
where $t=1,2, ... ,n_b - k +1$, the $k^{th}$ LSTM works as the following: 
\begin{equation}
\begin{gathered}
f^k_t = \sigma (W^k_f[z^0_k[t], h_{t-1}] + b^k_f)\\
i^k_t = \sigma (W^k_i[z^0_k[t], h_{t-1}] + b^k_i)\\
o^k_t = \sigma (W^k_o[z^0_k[t], h_{t-1}] + b^k_o)\\
c^k_t = f^k_t * c^k_{t-1} + i^k_t * tanh(W^k_c [z^0_k[t], h^k_{t-1}] + b^k_c)\\
h^k_t = o^k_t * tanh(c^k_t)
\end{gathered}
\end{equation}
where $\sigma()$ is the hard sigmoid activation function,$f^k_t$, $i^k_t$, and $o^k_t$ are the values of the forget gate, input gate, and output gate. $c^k_t$ is the state. $W^k_f, W^k_i, W^k_o, W^k_c$ 
and $b^k_f, b^k_i, b^k_o, b^k_c$ are the weights and biases for the corresponding gates, $h^k_t$ is the output, and $[]$ is the concatenation operator. 
The computation of the second layer  $z^{1}$ is thus 
\begin{equation}
z^{1}_k = LSTM_k(z^{0}_k) = h^{k}_{n_b - k +1},\quad k = 1,2,3,...,l
\end{equation}

At the third layer of MLP, all outputs of LTSMs are concatenated together  and fed into the layer to generate the final relevance decision.  
The computation of the third layer $s$ is  
$
s = MLP([z^{1}_{1}, ... z^{1}_{l}])
$. 

To summarize,  the overall architecture (see Figure \ref{fig:network}) of   DeepTileBars is: 
\begin{equation} \small
z^{0}_k = CNN_k(I), \quad k = 1,2,3,...,l
\end{equation}
\begin{equation}  \small
z^{1}_k = LSTM_k(z^{0}_k),\quad k = 1,2,3,...,l
\end{equation}
\begin{equation}  \small
s = MLP([z^{1}_{1},z^{1}_{2},... ,z^{1}_{k}, ... ,z^{1}_{l}])
\label{formula:ranking_fn}
\end{equation} 

 We use a state-of-the-art Stochastic Gradient Descent (SGD) optimizer, Adam \cite{kingma2014adam}, to optimize the  network. We  also adopt  the $L2$ regularization (at the  first layer) and early stopping 
 to avoid overfitting.
The document ranking scores are obtained by a pairwise ranking loss function as in RankNet \cite{burges2005learning}. It maximizes the difference between the relevant documents and the irrelevant documents:   $J(\Theta) = \sum_{(q, d^+, d^-)} - \log \frac{1}{1 + e^{-(s(d^+, \Theta) - s(d^-, \Theta))}}$, 
where $s$ is neural network output (Eq. \ref{formula:ranking_fn}), $\Theta$ are the network parameters, and $q$, $d^+$, and $d^-$  are  query,   relevant document and irrelevant document, respectively.

\section{Experiment}\label{sec:exp}



We evaluate  DeepTileBars' effectiveness on standard  testbeds, including TREC Web Tracks and LETOR 4.0. We compare the performance of DeepTileBars with both traditional IR approaches and  the state-of-the-art Neu-IR models.

\subsection{Dataset and Metrics}

We conduct experiments on the TREC 2010-2012 Web Track Ad-hoc 
tasks \cite{trec2012}.\footnote{TREC 2009 also had the Web Track. However, it used  relevance scales and  metrics quite different from the other years. For fairness and consistency, our experiments exclude year 2009.}   
There are 50 queries each year. We combine the three years' data as a single dataset because the annotated data from any single year is not enough to train a deep model.  In total, there are 150 queries and 38,948 judged documents. Our experiment is conducted on ClueWeb09 Category B, which contains more than 50 million English webpages (231 Gigabytes in size).\footnote{\url{http://lemurproject.org/clueweb09/}.} We implement a 10-fold cross-validation for fair evaluation. The official metrics used in TREC 2010-2012 Web Track ad-hoc tasks include 
Expected Reciprocal Rank (ERR)@20 \cite{chapelle2009expected}, normalized Discounted Cumulative Gain (nDCG)@20 \cite{jarvelin2002cumulated} and Precision (P)@20.  ERR and nDCG handle graded relevance judgments and Precision handles binary relevance judgements.

We also test our full model on the most recent MQ2008 dataset for LETOR 4.0. LETOR 4.0 is a common benchmark used by Neu-IR models. 
 LETOR MQ2008  contains 784 queries and 15,211 annotated documents. The official metrics used in LETOR includes nDCG and Precision at different cutoff positions.  We report results on queries that have at least  one relevant documents in the judgment set.


\subsection{Systems to Compare}

\begin{itemize}
	\item Traditional IR models: \textbf{BM25} \cite{robertson2009probabilistic} and \textbf{LM} \cite{zhai2017study}. Both are highly effective IR models. 

	\item TREC Best Runs: TREC Best is not a single system. Instead, it is a combination of best systems per year per metric, including the best systems in 2010 \cite{elsayed2010umd,dinccer2010irra}, 2011 \cite{boytsov2011evaluating} and 2012 \cite{al2012query}. We believe they represent the best  performance of TREC submissions from 2010 to 2012.



	\item Neu-IR models: 
	\textbf{DRMM}, \textbf{MatchPyramid}, \textbf{DeepRank}, \textbf{HiNT}, \textbf{DUET}.\footnote{We did not run DUET on TREC Web Tracks because the dataset is too  small to train a very deep model like DUET that requires a large amount of training data. }

\item 
Variations of the proposed model: 
\textbf{DeepTileBars ($n_{q} \times x $)} that uses a CNN with a single kernel size  $n_{q} \times x$; \textbf{DeepTileBars (w2w, all kernels)}, which uses all kernels  with word-to-word matching; and \textbf{DeepTileBars(w2s, all kernels)}, our full model, using word-to-segment matching and all kernels.

\end{itemize}

\subsection{Parameter Settings}



For the TextTiling algorithm, we set $\alpha$ to 20 and $\beta$ to 6, as recommended by \citeauthor{texttiling}. 
For the query-document interaction matrix, the parameters are  slight differences between  the two datasets.  In TREC Web, $n_{q}=5$ and In LETOR $n_{q}=9$.  $n_q$ is decided by the longest query after removing stopwords. For both datasets,   $n_{b}=30$. This is because more than 90\% documents in TREC Web and more than 80\% documents in LETOR contain no more than $30$ segments.


For the  DeepTileBars algorithm, we set $l=10$ for both datasets. In TREC Web Track dataset, the number of filters of CNN with same kernel size and the number of units in each LSTM are both set to $3$; while in LETOR, this number is  set to 9. The  MLP contains two hidden layers, with $32$ and $16$ units for TREC Web, and $128$ and $16$ units for LETOR.


On the TREC Web dataset, we re-implement DRMM, MatchPyramid, DeepRank, HiNT by following the configurations in their original papers. On LETOR, we  include the reported results in their  publications in Table \ref{tab:exp_mq2008} without repeating the experiments.

\subsection{Results}

\begin{table}
\centering
\small
\scalebox{1}{
\begin{tabular}{c|ccc}
\toprule
Run 	&	err@20	& ndcg@20 & p@20 \\
\midrule
TREC-Best &	\textbf{0.188}	& \textbf{0.236}	&	0.382 \\
\hline
BM25 	& 0.102 &	0.137 	& 0.253 \\
LM 		& 0.118	&	0.166	& 0.297 \\
\hline
DRMM & 0.127 & 0.184& 0.346 \\
MatchPyramid  & 0.113 & 0.125 & 0.228 \\
DeepRank  & 0.127 & 0.134 & 0.224 \\
HiNT  & 0.157 & 0.205 & 0.322 \\
\hline
DeepTileBars ( $n_{q} \times 1$) & 0.140 & 0.207 & 0.368\\
DeepTileBars ($n_{q} \times 3$) & 0.150 & 0.212 & 0.369\\
DeepTileBars ($n_{q} \times 5$) & 0.146 & 0.211 & 0.371\\
DeepTileBars ($ n_{q} \times 7$) & 0.142 & 0.207 & 0.366\\
DeepTileBars ($n_{q} \times 9$) & 0.147 & 0.213 & 0.372\\
\shortstack{DeepTileBars (w2w, all kernels)}&  0.110 & 0.123 & 0.248\\
\shortstack{DeepTileBars (w2s, all kernels)}&	0.168	& 0.229		& \textbf{0.384} \\
\bottomrule
\end{tabular}}
\caption{TREC 2010-2012 Web Track Ad-hoc Tasks. } 
\label{tab:exp}
\end{table}

\begin{table} 
\small
\scalebox{1}{
\begin{tabular}{c|p{0.5cm}p{0.5cm}p{0.7cm}p{0.9cm}l}
\toprule
Run 	& p@5	& p@10 &	ndcg@5 &ndcg@10 & Reported by\\
\midrule
BM25		& 0.337 &   0.245 & 0.461	& 0.220 & \cite{letor}\\
LM 		& 0.323	&   0.236 & 0.441	& 0.206 & \cite{letor}\\
\hline
DRMM	& 0.337	&	0.242 & 0.466	& 0.219 & \cite{pang2017deeprank}\\
MatchPyramid & 0.329 & 0.239 & 0.442 & 0.211 & \cite{pang2017deeprank}\\
DeepRank & 0.359 & 0.252 & 0.496 & 0.240 & \cite{pang2017deeprank}\\
Duet	& 0.341	&	0.240 & 0.471	& 0.216 & \cite{hint2018}\\
HiNT 	& 0.367	& 	0.255 &  0.501	& 0.244 &\cite{hint2018}\\
\hline
DeepTileBars & \textbf{0.427} & \textbf{0.320} & \textbf{0.553} & \textbf{0.256}&\\
\bottomrule
\end{tabular}
}
\caption{LETOR-MQ2008.}
\label{tab:exp_mq2008} 
\end{table}

Tables \ref{tab:exp} and  \ref{tab:exp_mq2008} report our experimental results for the TREC Web Tracks and the LETOR MQ2008 datasets. Official metrics are reported here. It can be found that in the TREC Web Track dataset, DRMM, HiNT and our model DeepTileBars outperform traditional IR approaches. While other neural IR approaches, MatchPyramid, DeepRank do not  perform  as well on certain metrics. On the LETOR dataset, all the Neu-IR approaches achieve better performance than traditional methods.
It indicates that properly designed deep neural networks could  improve upon traditional approaches. We also notice that with different network architectures and different input formulations, the Neu-IR models achieve varied gains. It would be worthwhile exploring which architecture is a better fit for ad-hoc retrieval.



It is exciting to see that our model, DeepTileBars, outperforms other state-of-the-art neural IR systems in TREC Web Tracks and MQ2008. 
We think the gains come from the topical segmentation of the texts and the bagging of multiple CNNs with different kernel sizes. Word-to-segment matching provides relevance signals at the topic level. CNNs with different kernel sizes  allow to evaluate document relevance at all  granularities. 


To investigate how DeepTileBars works internally, we experiment with a few variants of DeepTileBars with only one kernel. It can be found that adjusting the kernel size can only bring marginal improvement while the combination of all kernels boosts the retrieval performance. 

When changing word-to-segment matching back to word-to-word matching, we observe a huge performance decrease. This confirms our claim that  word-to-word matching is less desirable than word-to-segment matching in terms of finding relevance patterns. 


However, we are still left behind by the TREC Best runs in some metrics. Some TREC Best runs used sophisticated term weighting methods without any deep learning \cite{dinccer2010irra,al2012query}. Others used shallow neural networks or linear regression but with  abundant  feature engineering \cite{elsayed2010umd,boytsov2011evaluating}. We look forward to the day when Neu-IR could catch up with the TREC best systems.

\section{Related Work} \label{sec:related}

There are only a few pieces of research that share similar intention with ours, i.e. visualizing the relevance signals in  a document for deep learning. Works that are the  closest to ours are MatchPyramid,  HiNT, and ViP \cite{fan2017learning}. 

MatchPyramid was proposed for text matching tasks, including paraphrase identification and paper citation matching. By plotting the similarity between two sentences in an $n \times m$ matrix, where $n$ and $m$ are the lengths of two sentences respectively, deep neural networks were used to find the matching patterns between sentences with this image-like visualization. In MatchPyramid, the matching is done at the word level and experiments show it is less effective. 




The more recent Neu-IR model, HiNT,  adopted a similar idea to ours to perform passage-level retrieval.  One major difference between HiNT and DeepTileBars is how they split the documents.  HiNT splits documents by fixed sized passages whereas DeepTileBars splits them based on topic changes. The passages in HiNT are in fact  more similar to our token sequences, which do not represent topical structure. As a result, the highest semantic level that HiNT is able to examine is similar to our segments. Levels higher than segments are not modeled in HiNT. In this sense, HiNT is not a true hierarchical Neu-IR model; instead, it is a segment-level only model. Meanwhile, although HiNT used a much more complex neural method, with a light-weighted architecture, DeepTileBars achieves better retrieval effectiveness in our experiments.

ViP also proposed to take advantage of visual features in a document. Instead of using the interaction matrix  as the input image, ViP directly used a webpage's snapshot as so. ViP's experiments showed that even query-independent visual features would be able to improve the retrieval effectiveness. ViP's semantic units, such as the webpage sections and multimedia components,  are similar to  our hierarchical topics  higher than  the segments. In this sense, they are the most similar work to ours. 
 However, we did not compare to their work in this paper due to our focus on texts.

Research from the field of Natural Language Processing has adopted similar designs of using multiple CNN kernel sizes. For example,  \cite{kim2014convolutional} used that for sentence classification. Their input was a $d\times m$ interaction matrix, where $d$ was the dimension of the term vectors and $m$ was the number of words in a sentence. They encoded a sentence using multiple CNNs with kernel sizes $d \times n$, where $n$ was the size of an $n$-gram and obviously  could vary. While our  use of multiple kernel sizes is driven by the attempt to  fuse relevance signals at multiple topical granularities, their modern way of representing $n$-grams (with different n) yielded a similar design to ours.  

\section{Conclusion}
\label{sec:conclusion}

 
In this paper, we propose DeepTileBars, a Neu-IR model inspired by classical work in search engine visualization. The main contribution includes  (1) word-to-segment matching and (2) bagging of different sized CNNs. Experiments show that our approach outperforms the state-of-the-art Neu-IR models.  One exciting property about our work is that it segments a document roughly by topics. We think  these topical segments are  more meaningful units than segments of fixed lengths and natural paragraphs (which do not necessarily respect topical boundaries). Moreover, with multiple different sized kernels,  a hierarchical modeling of  document structure is practically enabled and   probably contributes to the effectiveness of our approach.



\section{Acknowledgement}
This research was supported by NSF CAREER Award IIS-1453721. Any opinions, findings, conclusions, or recommendations expressed in this paper are of the authors, and do not necessarily reflect those of the sponsor.

\bibliography{reference}
\bibliographystyle{aaai}

\end{document}